\documentclass{article}
\usepackage{graphicx} 
\usepackage{xcolor}
\usepackage{multirow}
\usepackage{booktabs}
\usepackage{tabularx}
\usepackage{hyperref}
\usepackage{amsmath}
\usepackage{mathtools}
\usepackage{bbm}
\usepackage{amsfonts}
\usepackage{bm}
\usepackage{adjustbox}
\usepackage{tikz}
\usepackage{pgfplots}
\pgfplotsset{compat=1.18}
\usepackage[backend=biber, style=apa]{biblatex}
\AtEveryBibitem{\clearfield{month}}
\AtEveryBibitem{%
  \ifentrytype{book}{\clearfield{url}}{}
  \ifentrytype{article}{\clearfield{url}}{}
}
\NewDocumentCommand{\citep}{O{} O{} m}{%
  \parencite[#1][#2]{#3}%
}
\NewDocumentCommand{\citet}{O{} O{} m}{%
\citeauthor{#3}~(\citeyear{#3})%
}
\usetikzlibrary{
shapes,
arrows, shapes.geometric,
shapes.arrows,
positioning,
decorations.pathreplacing,
angles,quotes,
fit,
calc, 
shapes.symbols} 
\NewDocumentCommand\Cycle{O{} m m m O{} m m}{%
  \draw[#1,<->, very thick, >=stealth'](#2.{#3+asin(#6/(#4*1.41))}) arc (180+#3-45:180+#3-45-#7:#6/2) #5;
}

\def\lambdas{1.236}
\def\lambdabpm{1}
\def\IntStress{-40.782}
\def\rhoStress{0.895}
\def\thetaStressEins{75.735}
\def\thetaStressZwei{28.06}
\def\vbpmNull{24.9}
\def\vbpm{6.075}
\def\vStress{86.329}
\def\IntFF{89.046}
\def\IntSeins{-0.431}
\def\IntSzwei{0.633}
\def\vI{164.518}
\def\vSeins{0.121}
\def\vSzwei{0.114}
\def\covISeins{-1.428}
\def\covISzwei{2.505}
\def\covSeinsSzwei{-0.023}
\def\ARFF{0.658}
\def\vFFNull{82.646}
\def\vFF{19.393}

\title{Time-dependent structural equation modeling of fans' football fever using activity tracking data during the 2025 DFB Cup final}
\author{Jonas Bauer, Christiane Fuchs,
and Tamara Schamberger}
\date{}
\begin{document}
\maketitle
\begin{abstract}
Football fans frequently exhibit pronounced emotional and physiological reactions during high‑stakes matches. However, the temporal dynamics of this \emph{football fever} are rarely modeled as a latent process. Using intensive longitudinal data from Arminia Bielefeld supporters who wore smartwatches during the 2025 German Football Association (DFB) Cup final, we investigate how football fever unfolds. The devices recorded heart rate, stress level, and related indicators in short intervals, allowing us to construct a latent variable for football fever and model its dynamics. We specify a time‑dependent structural equation model with latent growth components and autoregressive effects to capture both overall trends and short‑term carry‑over effects in fans’ physiological responses. Results are aggregated across multiple imputations of missing measurements. Model fit is evaluated using adjustments for the high data dimensionality. The results show that football fever follows a V‑shaped trajectory: high at kick‑off, followed by a steady decline until the renewed arousal in the second half, with substantial between‑fan heterogeneity in both baseline level and temporal dynamics. Our findings demonstrate that football fever can be adequately represented as a latent variable using structural equation modeling and reflected by wearable technology data. This highlights the importance of accounting for temporal dependence when studying dynamic emotional phenomena, e.\,g., in sports spectatorship.
\end{abstract}

\clearpage
\section{Introduction}
Football is arguably the most popular sport worldwide and a rapidly growing economic industry with trans-national importance \citep{VeliciaMartin2020HowTeams}.
More than four billion fans per season watch the UEFA Champions League \citep{Kennedy2012FootballEurope}, two billion fans engaged in the FIFA Women's World Cup 2023 in Australia \& New Zealand \citep{fifa2023audience}, and even five billion fans in the FIFA Men's World Cup 2022 in Qatar \citep{fifa2022audience}.
These billions of fans drive the football industry through regular consumption and shaping the live atmosphere \citep{Abosag2012ExaminingClubs}.
The European football market, for instance, is steadily growing \citep{Rohde2017TheFootball}, reaching a market value of 38 billion euros in the season 2023/24 \citep{deloitte2025}.

Given the popularity of football and fans' central role in fueling this industry, it is critical to understand fan behavior and underlying mechanisms. 
A vast literature reflects this scientific interest.  
Surveys have been used, for example, to investigate consumption motives \citep[e.\,g.][]{Cho2019NostalgiaTourism, Correia2007AnFootball, Stander2016SeeAfrica}, emotions \citep[e.\,g.][]{Abosag2012ExaminingClubs, VeliciaMartin2020HowTeams}, fan engagement \citep{Vale2018SocialFacebook, Chen2025SportIntensity} and loyalty \citep{Kim2021ExaminingFootball}.
In addition, the impact of sports consumption on fans' health has been researched qualitatively based on a diary study and interviews \citep{Pringle2004CanMen}. 
Moreover, \citet{vanderMeij2012TestosteroneFinal} combined survey data of fans with their hormonal reactions at three time points (right before, in the halftime break and right after the game), examining fans' anticipation of the outcome on their social esteem.
\par
Still, these studies show two limitations:
First, they are based on surveys asking about past experiences (e.\,g., ``How frustrating/stressful was watching the final for you?"\ from \citet{vanderMeij2012TestosteroneFinal}), which are often prone to bias \citep{Levine2002SourcesEmotions}.
The reason for such bias is that participants' answers are not based on time-invariant pieces of information, but memories of past emotions, which are reconstructed in light of current emotions and coping efforts \citep{Levine2002SourcesEmotions}.
Second, the majority of the studies use cross-sectional data, i.\,e., snapshots of a particular point in time, despite repeated calls for longitudinal designs that can incorporate change over time \citep[e.\,g.,][]{Santos2019ExaminingSites, Santos2022ConsumerReview, DeBeer2016ForOutcomes.}.
Hence, they do not incorporate temporal dynamics.
Studies with repeated measurement time points have rather low resolutions, allowing for differentiation (e.\,g., differences between halftimes), but neither permit explicit modeling of match dynamics nor evaluating of critical events such as goals scored.
\par
In contrast, the advent of wearable technology facilitates longitudinal data collection with much higher temporal resolution. 
Accordingly, in previous work \citep{Adam2026MeasuringTechnology}, we measured and described vital parameters among almost 200 football fans during a high-stakes match, the 2025 Cup final of the German Football Association (DFB) between DSC Arminia Bielefeld and VfB Stuttgart.
We found fans' heart rate and stress level elevated throughout the match and pronounced spikes, especially following goals scored.
Further, we hypothesized these descriptive findings might be driven by an underlying \emph{football fever} among fans, yet a model-based examination of this relationship remains to be carried out.
In the present article, we explicitly model the unobservable football fever as a theoretical construct, manifested through observable physiological variables.
We further examine its temporal dynamics over the course of the match, aiming to provide meaningful insights, for instance, concerning targeted real-time and fan-specific communication, live content production, or social media strategies. 
Model assessment is based on procedures specifically accounting for missing measurements \citep{Chan2022MultipleTests} and the high dimensionality of intensive longitudinal data \citep{Yuan2015EmpiricalVariables}.

%
\par
%
%
To model the theoretical construct, we use structural equation modeling \citep[SEM,][]{Bollen1989}.
While SEM can incorporate three different types of constructs, namely composites, causal-formative constructs, and common factors \citep{Henseler2021,Bauer2025MisspecificationsComposites, Schamberger2025StructuralCompositesArxiv}, we specifically model football fever as a common factor, i.\,e., the underlying common cause for the physiological measurements \citep{Joreskog1970a}.
As the wearable technology data is a multivariate time series, the measurements are neither independently nor identically distributed, but correlated over time.
We account for the temporal structure of football fever using latent state-trait theory to decompose its variance into a stable, time-invariant trait and a transient, time-dependent state \citep[e.\,g.,][]{Norget2025EstimatingTutorial, Steyer2015ARevised}.
Complementarily, we parameterize fan-specific trajectories via a latent growth curve, i.\,e., we incorporate random effects in the trend \citep[e.\,g.,][]{Mayer2012AComponents, Bollen2005LatentPerspective}.

The following section details the data and its pre-processing.
Section~\ref{sec:analysis} formally introduces our time-dependent structural equation model, followed by the empirical results.
The article closes with a discussion of our findings in Section \ref{sec:discussion} and concluding remarks in Section \ref{sec:conclusion}.

\section{The 2025 DFB Cup final data} \label{data}
On May 24, 2025, the DFB Cup final between third-division club DSC Arminia Bielefeld and first-division club VfB Stuttgart took place in Berlin, Germany. 
For Arminia Bielefeld, this event was historic in many ways: It was the first time in the club's history to reach the final of this tournament, and it became the first time ever a third-division club scored a goal in the Cup final.
Accordingly, this match was exceptional for the football fans of Arminia Bielefeld.
In \citet{Adam2026MeasuringTechnology}, we 
collected high-resolution smartwatch data to evaluate the effect of such a high-stakes match on Arminia Bielefeld fans.
Our descriptive study showed that fans' experience elevated heart rates and stress levels during the game compared to regular days with pronounced spikes at kick-off and decisive moments, indicating a potential common cause, i.\,e., football fever.
In the present article, we investigate this hypothesis by modeling and examining the theoretical construct football fever and its relations to fans' heart rates and stress levels in greater detail. 
We next provide a concise overview of the main data characteristics, focusing on key variables of the subsequent analysis and the time horizon of the Cup final.
We keep this overview brief, as our primary aim here is to familiarize the reader with the most relevant aspects of the data (including the issue of missing observations and our imputation strategy) rather than to present a full exposition of the dataset.
A more detailed description of the complete observation period and further information on the study design can be found in \citet{Adam2026MeasuringTechnology}.
\par
\begin{figure}[ht]
    \centering
    \includegraphics[width=0.95\linewidth]{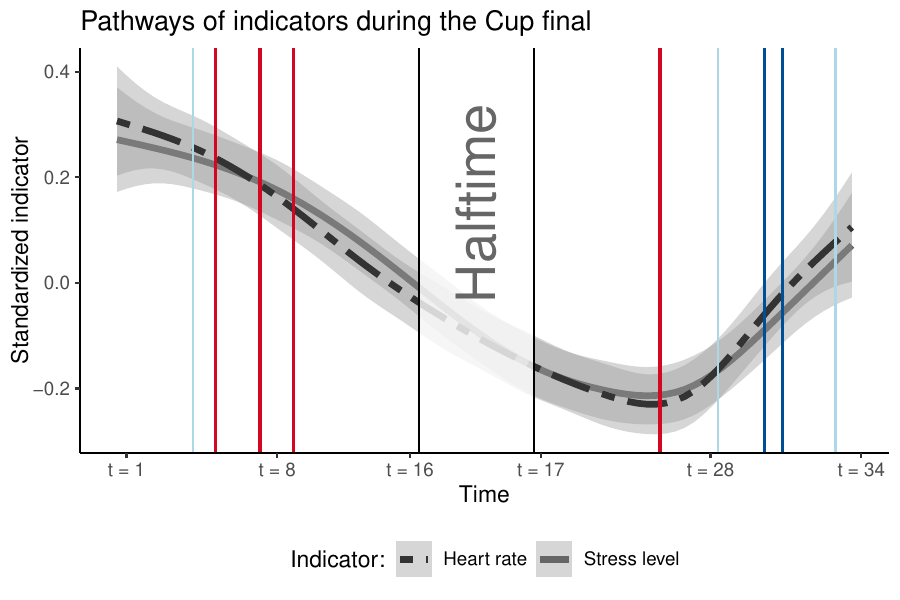}
    \caption{Standardized averages of heart rate and stress level, smoothed over match time~$t$ using a generalized additive Gaussian model. 
    Vertical lines indicate key match events: Goals by VfB Stuttgart (red) and DSC Arminia Bielefeld (blue), as well as major unsuccessful scoring opportunities for Bielefeld (light blue). The faded interval in the center corresponds to the halftime break, separating measurements of the first half ($t=1,\dots, 16$) from the second half ($t=17,\dots,34$).}
    \label{fig:smoothed_inds}
\end{figure}
The employed smartwatches from the manufacturer Garmin measure the heart rate by an optical sensor in beats per minute 
\citep{FitabaseGarminDataDictionary}.
The stress level is based on a combination of heart rate, its variability between consecutive heartbeats, and additional personal factors that distinguish individual stress from relaxation states \citep{firstbeat2014}.
The employed devices report stress values within intervals of three minutes on a user-specific scale from 0 (no stress) to 100 (maximal stress).
To match the measurement intervals of both variables, we average the reported heart rates of each fan within those three minute intervals.
After this aggregation, the data comprises 34 equidistant measurement time points over the course of the Cup final, with $t=1,2,\dots, 16$ representing the first half and $t=17,18, \dots, 34$ the second half of the 97 match minutes.
Figure~\ref{fig:smoothed_inds} visualizes the pathways of average heart rate and stress level among Arminia Bielefeld fans.
Both variables have been standardized and the curves are derived from generalized additive models with integrated smoothness estimation after standardization.
The average fans' heart rate peaked at kick-off and remained elevated during the first 20 minutes of the match.
Over the course of the match, the average heart rate decreases monotonically until it rises again over the last 30 minutes of the match.
While the steady drop is likely due to the early goals of the opponent, VfB Stuttgart, the late rally appears to be dominated by the goals and chances of Arminia Bielefeld.
Furthermore, Figure~\ref{fig:smoothed_inds} reveals a synchronized temporal pattern between the heart rates of fans and their stress levels, 
which leads to a strong empirical correlation of 0.77 between them. 
\par
%
For both heart rate and stress level, some measurements are missing.
Among the 194 participants with heart rate measurements during the Cup final, there are 8 fans without any stress level records.
We excluded these fans from our study.
Moreover, we removed 49 additional fans for whom more than 50\% of stress level measurements during the match are missing.
The remaining 137 participants still have 25\% missing values in their stress level on average.
These missing values can have various reasons, e.\,g., sensor failure, transfer error, or recording error.
Additionally, the smartwatches do not record stress levels during high physical activity because exertion itself causes increases in heart rate variability.
Consequently, the missingness coincides with other activity-based variables, such as increased motion intensity, more steps taken, and a higher calorie burn.
\par
Out of the strategies to deal with missing stress levels, list-wise deletion of fans would discard excessive information, and model-based imputation within SEM likely produces non-positive definite sample variance-covariance matrices due to strong dependencies between consecutive time points.
Therefore, we predict the missing stress levels conditional on the available activity-based measurements. 
In particular, this multivariate imputation approach uses the full-conditional distribution of a Gibbs sampler and often yields unbiased parameter estimates, even if this conditional distribution had been misspecified \citep{VanBuuren2006FullyImputation}.
Random forests are a frequently used and versatile approach to estimate such conditional distributions \citep{Cevid2022DistributionalRegression}.
Consequently, we estimate the full-conditional distribution with a random forest, which is then used to impute missing stress levels.
Since imputed values are random draws, one should account for their uncertainty by repeating this procedure multiple times.
Hence, we sample ten imputed datasets using the multivariate multiple imputation strategy implemented in the R package \texttt{mice} \citep{VanBuuren2011mice}.
As the imputation technique affects the subsequent parameter estimation and other results of our work, we provide the results based on other distributional assumptions in the Appendix \ref{app:result_comparison}.
%
%
\section{Statistical modeling of the dynamics of football fever} \label{sec:analysis}

The synchronicity between heart rate and stress level, observed in Figure~\ref{fig:smoothed_inds}, suggests an underlying common factor, football fever, which we model within a structural equation model.

\subsection{A structural equation model with time-dependencies} \label{sec:model}
\begin{figure}[!htbp]
    \centering
    \adjustbox{rotate=-90}{
    \begin{minipage}{1.3\linewidth}
    \input{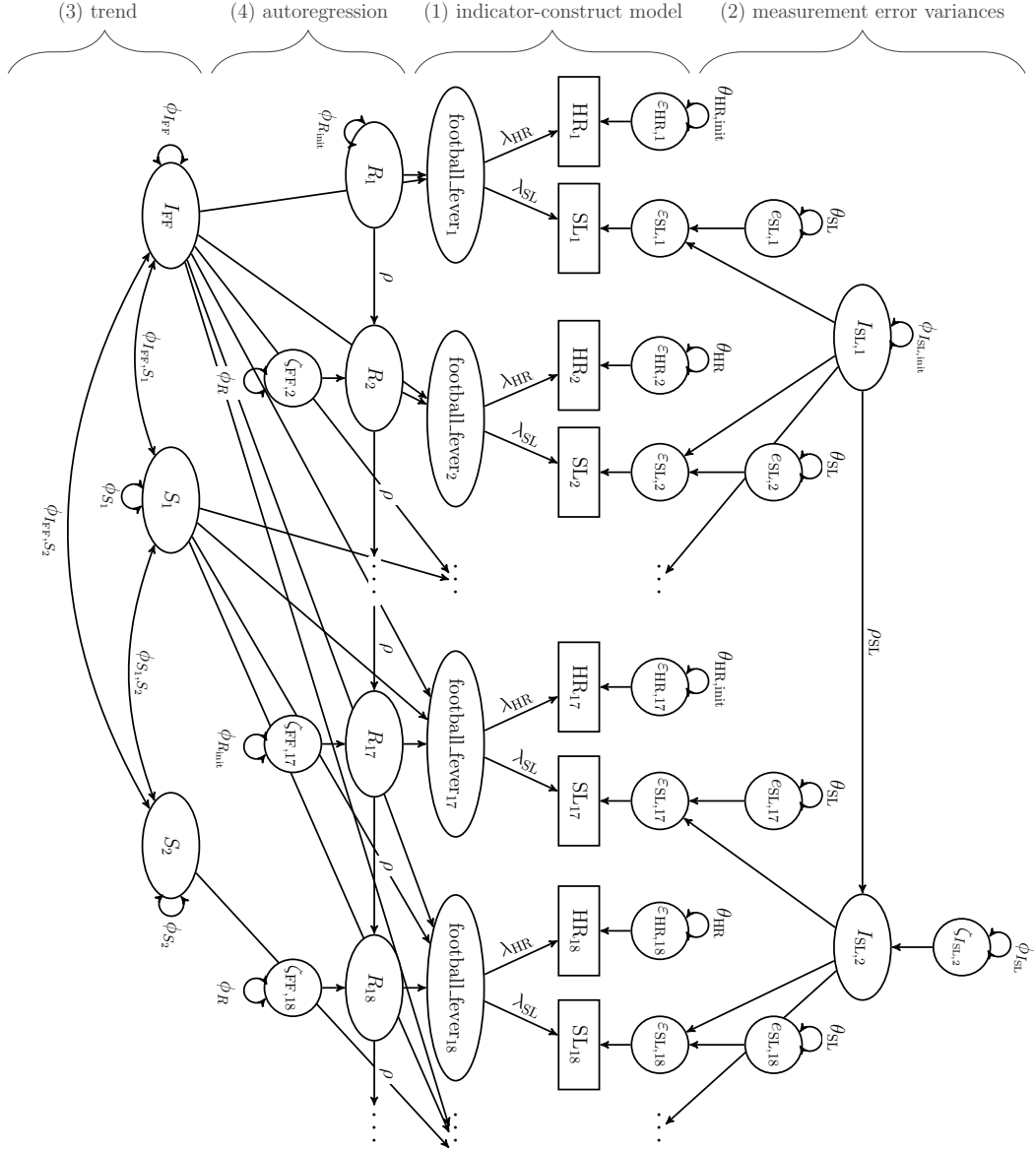}
    \end{minipage}
    }
    \caption{Our structural equation model for the dynamics of football fever. Coefficients of paths without a parameter or value are fixed to 1, except those from $S_1$ and $S_2$ onto $\text{football\_fever}_{t}$. These coefficients are fixed to $t-7$ for the paths $S_1 \to \text{football\_fever}_{t}, t>7$ and to $t-25$ for the paths $S_2 \to \text{football\_fever}_{t}, t>25$, respectively. The intercepts of $\text{SL}_1, \dots, \text{SL}_{34}, I_{\text{SL},1}, I_{\text{SL},2}, I_{\text{FF}}, S_1$, and $S_2$ are freely estimated but omitted here for readability.}
    \label{fig:SEMtime}
\end{figure}

Figure \ref{fig:SEMtime} illustrates our structural equation model, displaying 4 of the 34 time points. 
The first two time points represent the beginning of the match and the other two time points the beginning of the second halftime.
The curly brackets in the Figure divide model variables into four categories: (1)~the indicator-construct model, describing the relationship between indicators and football fever, (2)~the measurement error variances, (3)~the trend of football fever, and (4)~the autoregression between consecutive time points in football fever.
Next we describe each part of the model in more detail.
\par 
In the indicator-construct model, we assume the common factor football fever ($\text{football\_fever}_t$) to be reflected by the two observed indicators, heart rate ($\text{HR}_t$) and stress level ($\text{SL}_t$), at time~$t$ as follows:
\begin{alignat}{3}
    \text{HR}_t      &= &&\lambda_{\text{HR}} &&\cdot\text{football\_fever}_t + \varepsilon_{\text{HR}, t} \label{eq:icmbpm} \\
    \text{SL}_t   &= \mu_\text{SL} + &&\lambda_\text{SL} &&\cdot \text{football\_fever}_t + \varepsilon_{\text{SL}, t}. \label{eq:icmstress}
\end{alignat}
The strength of the effect of football fever on the two indicators is parameterized by the factor loadings $\lambda_{\text{HR}}$ and $\lambda_{\text{SL}}$ in Equations \ref{eq:icmbpm} and \ref{eq:icmstress}.
We assume changes in football fever to affect the set of indicators in exactly the same way, regardless of the time point at which they occur during the match. Thus, we specify both factor loadings as constant over time.
We further specify heart rate as marker indicator by fixing its factor loading in Equation \ref{eq:icmbpm} to one, providing a scale to the common factor \citep[e.\,g.][]{Kline2023}.
\par
Measurement errors have an expectation of zero, but their variances capture fans' variability at each time point. 
We assume these variances are affected by both individual (i.\,e.\ fan-specific) and temporal (i.\,e.\ time point specific) factors.
For instance, fans might exhibit a greater heterogeneity in their heart rate at the start of each half (i.\,e., $t=1$ and $t=17$) due to more diverse activities compared to other time points; while some fans are resting (e.\,g., already seated and awaiting the starting whistle), others are physically more active (e.\,g., returning from a toilet break or purchasing food and beverages).
To account for these individual factors, we estimate the measurement error variances for the heart rate at both halftime starts using a single free parameter, denoted by~$\theta_{\text{HR},\text{init}}$.
At all other time points, they are assumed to be time-invariant and denoted by $\theta_{\text{HR}}$, yielding the following conditional measurement error variances for heart rate:
\begin{align*}
    \mathbb{V}(\text{HR}_t|\text{football\_fever}_t)  =  &\begin{cases}
        \theta_{\text{HR},\text{init}} & \text{for }t \in \{1,17\} \\
        \theta_{\text{HR}} & \text{elsewhere}
    \end{cases} 
\end{align*}
Fans' individual stress responses to match events might vary beyond what can be attributed to football fever. 
We thus distinguish two types of measurement error variances in stress levels: Within-time-point variability and between-time-point variability.
The individual differences of fans within time points are quantified by variables~$e_{\text{SL},t}$ with time-invariant variance~$\theta_\text{SL}$.
For the between-time-point variability in fans' stress levels we include random intercepts~$I_{\text{SL},1}$ and $I_{\text{SL},2}$ associated with all measurement errors of the stress level for the first and second halftime, respectively.
Consequently, $\theta_\text{SL}$ quantifies the stable variability between fans throughout the match, and variances~$\phi_{I_{\text{SL},\text{init}}}$ and~$\phi_{I_{\text{SL},2}}$ of the random intercepts capture differences between time points:
\begin{align*}
    \mathbb{V}(\text{SL}_t|\text{football\_fever}_t) &= \theta_{\text{SL}} + \begin{cases}
        \phi_{I_{\text{SL},\text{init}}} & \text{for }t \in \{1,\dots, 16\} \\
        \rho^2_{\text{SL}}\cdot \phi_{I_{\text{SL},\text{init}}} + \phi_{I_{\text{SL},2}} &  \text{for }t \in \{17,\dots, 34\} ,
    \end{cases}
\end{align*}
with $\rho_{\text{SL}}^2$ quantifying the carry-over effect from the first half's variance to the second half.
\par
The trend represents the long-term progression of football fever, modeling the joint pathway of heart rate and stress level over the course of the match. 
The pathways of both indicators (see Figure \ref{fig:smoothed_inds}) can be characterized by three phases: A plateau at the beginning, a steady decline following Stuttgart's early goals ($t>7$), and a steady incline during the late rally ($t>25$). 
Accordingly, we model the trend of football fever by three random variables as well: An overall intercept~$I_{\text{FF}}$, and two slopes~$S_1$ and~$S_2$.
The expected football fever is thus given as follows:
\begin{align}
\mathbb{E}\left(\text{football\_fever}_t\right) &=  \mu_{I_{\text{FF}}} +\begin{cases}
        0 & t\leq 7 \\
         (t-7)\cdot \mu_{S_1} & 7 < t \leq 25 \\
        18\cdot \mu_{S_1} + (t-25) \cdot \mu_{S_2} & t>25,
    \end{cases} \label{eq:trend}
\end{align}
where $\mu_{I_{\text{FF}}}$ denotes the expected value of the overall intercept~$I_\text{FF}$, $\mu_{S_1}$ the expected value of the first slope~$S_1$ (i.\,e., the linear change between $t=8$ and $t=24$), and $\mu_{S_2}$ the expected value of the second slope~$S_2$ (i.\,e., the linear change starting from $t=25$).
Moreover, fans likely differ in their initial football fever and their reactions to match events. 
To account for this individual heterogeneity, we allow all trend variables to (co-)vary freely by estimating their variances~$\phi_{I_\text{FF}}$, $\phi_{S_1}$, and $\phi_{S_2}$ as well as their covariances~$\phi_{I_\text{FF},S_1}$, $\phi_{I_\text{FF},S_2}$, and $\phi_{S_1,S_2}$.
Hence, the trend variables are 
correlated latent growth components \citep{Mayer2012AComponents}.
Each fan is thus associated with an individual $\text{trend}_t$, determined by a random draw of the trend variables~$I_\text{FF}, S_1$, and $S_2$.
These realizations replace the means in Equation \ref{eq:trend} to form that fans' conditional expected football fever.
\par
The autoregression is the fourth and last category of variables in our structural equation model. 
Fans are likely to differ not only in their expected football fever, but fluctuate around their individual trends, e.\,g., due to toilet breaks or being distracted.
Such fan-specific residuals around their respective growth curves might also carry-over from one time point to the next.
We incorporate this in our model by adding an autoregressive stochastic process~$\{R_t\}_t$:
\begin{align*}
    \text{football\_fever}_t  &= \text{trend}_t + R_t \\
R_t &= \rho R_{t-1} + \zeta_{\text{FF},t} \\
\mathbb{V}(\zeta_{\text{FF},t}) &= \phi_R
\end{align*}
for $t>1$.
We assume $R_t$ to have an expectation of zero.
Thus, it does not interfere with the trend of football fever; however, it increases the football fever's variance. 
The initial variance $\mathbb{V}(R_1)=\mathbb{V}(R_{17}|R_{16})=\phi_{R_\text{init}}$ accounts for the heterogeneity of fans at the start of the halftimes.
The disturbance terms~$\zeta_{\text{FF},t}$ quantify the added variability in each particular time point, i.\,e., the conditional variances $\mathbb{V}(R_t|R_{t-1}) = \phi_R, t>1$.
Since the process~$\{R_t\}_t$ is of order one, the carry-over  effect~$\rho$ determines how strongly the variance of the previous time point affects the current one.
For $|\rho|<1$, these carry-over effects fade away geometrically and the unconditional variance of $R_t$ is given by:  
$$\mathbb{V}(R_t)= \phi_{R_\text{init}}\cdot \rho^{2(t-1)} + \phi_R\cdot \sum_{j=0}^{t-2}\rho^{2j}.$$
\par
Overall, our structural equation model consists of 20 free parameters, which are spread across the indicator-construct model (2), the measurement error variances (6), the trend for football fever (9), and its autoregression (3).
This comprehensive framework enables us to distinguish and examine the different aspects of the football fever phenomenon, such as the average football fan in our sample, between-fan heterogeneity, as well as time-specific dynamics.
\subsection{Results} \label{sec:results}
In the following, we present the estimated parameters and the evaluated overall fit of our structural equation model.
All analyses were conducted in the statistical software \texttt{R} \citep{R4.5.2}, using the packages \texttt{lsttheory} \citep{lsttheory0.4-1} for model specification and \texttt{lavaan} \citep{lavaan0.6-20, Rosseel2012} for parameter estimation and model assessment\footnote{Source files are available online at \href{https://osf.io/bn4jh/overview}{https://osf.io/bn4jh/overview}.}. 
As we employ ten imputed datasets, we also obtain ten sets of model estimates.
While the point estimates can be pooled by taking their arithmetic means across all imputation datasets, standard errors are pooled by combining the within-imputation covariance matrices with the between-imputation covariance matrix \citep[see][]{Enders2010}.
We carry out this pooling using the methods implemented in the \texttt{R} package \texttt{lavaan.mi} \citep{lavaan.mi0.1-0}.
\par
Table \ref{tab:estimates} displays the pooled estimates of the 20 free parameters of our model, along with pooled standard errors and the p-values of t-tests on these pooled characteristics.
According to the t-tests, 17 of the 20 parameters are 
significantly different from zero at a significance level of $\alpha=0.05$.
Even when considering the Bonferroni correction \citep{Dunn1961MultipleMeans} for multiple testing, this pattern barely changes, though one additional parameter is rendered insignificant. 
However, the joint effect of the four parameters in question, assessed by a likelihood ratio test, justifies their inclusion in our model.
\par
In the indicator-construct model, the unstandardized factor loadings are constant throughout the match; that of stress level~$\lambda_\text{SL}$ is estimated as \lambdas, while that of~$\lambda_\text{HR}$ is fixed at 1.
In contrast, the standardized factor loadings vary over time due to different scales for football fever and the indicators, yielding an average of 0.80 for stress level and 0.98 for heart rate.
The squared standardized factor loadings indicate the proportion of variance in each indicator explained by football fever, averaging 96.6\% for heart rate and 64.4\% for stress level across the match.
Vice versa, the construct reliability of football fever (0.89 on average) quantifies how reliable heart rate and stress level measure this common factor.
During the Cup final, the expectations of heart rate and football fever are specified to be equal, indicating roughly 89 beats per minute for the average fan ($\hat\mu_{I_{\text{FF}}}$).
The average stress level at kick-off is $69$ (namely, $\hat\lambda_\text{SL}\cdot\hat\mu_{I_{\text{FF}}}+\hat\mu_{\text{SL}}$), lying on the upper edge of medium stress (between 51 and 75) on the manufacturer's scale.
\par
\begin{table}[ht]
    \centering
    \setlength{\tabcolsep}{3pt}
    \begin{tabularx}{0.825\textwidth}{llrlrrc}
    \toprule
\multicolumn{2}{c}{\textbf{Parameter\ }} & \multicolumn{2}{c}{\textbf{Estimate}} & \textbf{\ Std.\ error} & \textbf{\ p-value} & \textbf{$\alpha^{\text{Bonf}}$ corr.}\\
\midrule
    \parbox[t]{2mm}{\multirow{3}{*}{\rotatebox[origin=c]{90}{ICM}}} &  $\lambda_{\text{HR}}$ & \lambdabpm.000 & & & \\
    & $\lambda_\text{SL}$ & \lambdas & $^{***}$ & 0.032 & $<0.001$ & $^\clubsuit$ \\ 
    & $\mu_{\text{SL}}$ & \IntStress & $^{***}$ & 2.893 & $<0.001$ & $^\clubsuit$
\\ 
    \midrule
    \parbox[t]{2mm}{\multirow{6}{*}{\rotatebox[origin=c]{90}{Meas.\ error}}}
    & $\theta_{\text{HR},\text{init}}$ & \vbpmNull00 & $^{***}$ & 4.961 & $<0.001$ & $^\clubsuit$ \\ 
    & $\theta_{\text{HR}}$ & \vbpm & $^{***}$ & 0.693 & $<0.001$ & $^\clubsuit$ \\ 
    & $\theta_{\text{SL}}$ & \vStress & $^{***}$ & 2.510 & $<0.001$ & $^\clubsuit$ \\ 
    & $\rho_{\text{SL}}$ & \rhoStress & $^{***}$ & 0.076 & $<0.001$ & $^\clubsuit$ \\ 
    & $\phi_{I_{\text{SL},\text{init}}}$ & \thetaStressEins & $^{***}$ & 11.877 & $<0.001$ & $^\clubsuit$ \\ 
    & $\phi_{I_{\text{SL},2}}$ & \thetaStressZwei0 & $^{***}$ & 5.562 & $<0.001$ & $^\clubsuit$ \\ 
    \midrule
    \parbox[t]{2mm}{\multirow{9}{*}{\rotatebox[origin=c]{90}{Trend}}} & $\mu_{I_{\text{FF}}}$ & 	\IntFF & $^{***}$ & 1.377 & $<0.001$ & $^\clubsuit$ \\ 
    & $\mu_{S_{1}}$ & \IntSeins & $^{***}$ & 0.052 & $<0.001$ & $^\clubsuit$ \\ 
    & $\mu_{S_{2}}$ & \IntSzwei & $^{***}$ & 0.092 & $<0.001$ & $^\clubsuit$ \\ 
    & $\phi_{I_{\text{FF}}}$ & \vI & $^{***}$ & 26.194 & $<0.001$ & $^\clubsuit$ \\ 
    & $\phi_{S_1}$ & \vSeins & $^{**}$ & 0.040 & 0.002 & $^\clubsuit$ \\ 
    & $\phi_{S_2}$ & \vSzwei &  & 0.129 & 0.379 \\ 
    & $\phi_{I_{\text{FF}},S_1}$ & \covISeins &  & 0.751 & 0.057 \\ 
    & $\phi_{I_{\text{FF}},S_2}$ & \covISzwei & $^{*}$ & 1.272 & 0.049 \\ 
    & $\phi_{S_1,S_2}$ & \covSeinsSzwei & & 0.053 & 0.660 \\ 
    \midrule
    \parbox[t]{2mm}{\multirow{3}{*}{\rotatebox[origin=c]{90}{AR}}} & $\rho$ & \ARFF & $^{***}$ & 0.025 & $<0.001$ & $^\clubsuit$ \\ 
    & $\phi_{R_\text{init}}$ & \vFFNull & $^{***}$ & 11.042 & $<0.001$ & $^\clubsuit$ \\ 
    & $\phi_{R}$ & \vFF & $^{***}$ & 1.031 & $<0.001$ & $^\clubsuit$ \\ 
    \bottomrule    
    \end{tabularx}
    \caption{Estimation of our time-dependent structural equation model: Parameter labels, estimates and standard errors (both pooled across 10 imputed datasets), p-values. Significance codes with asterisks (***: $p < 0.001$, **: $p < 0.01$, *: $p < 0.05$; blanks indicate insignificance) refer to single post-hoc tests; significance after the Bonferroni correction for 20 tests (threshold $\alpha^{\text{Bonf}}=0.0025$) is indicated by black clubs ($^\clubsuit$).}
    \label{tab:estimates}
\end{table}
Fans differ substantially regarding their stress levels, with an average model-implied variance of 505.9, calculated as
\[
\frac{1}{34}\sum_{t=1}^{34}\hat{\mathbb{V}}(\text{SL}_t|\text{football\_fever}_t)+\hat\lambda_\text{SL}^2\cdot \hat{\mathbb{V}}(\text{football\_fever}_t),
\]
which can be further decomposed into between-fan variability, halftime-specific variability, and football fever variability, i.\,e., common factor variance.
In particular, 18\% of the model-implied variance of stress level are attributable to individual heterogeneity ($\hat\theta_\text{SL}/505.9$) and 15\% to time-specific heterogeneity ($\hat\phi_{I_\text{SL}}/505.9$).
In contrast, the average model-implied variance of heart rate is considerably smaller at 207.9 (that is, $\frac{1}{34}\sum_{t=1}^{34}\hat{\mathbb{V}}(\text{HR}_t|\text{football\_fever}_t)+\hat{\mathbb{V}}(\text{football\_fever}_t)$).
Individual heterogeneity accounts for only 3.6\% of that variance at regular time points ($\hat\theta_{\text{HR}}/207.9$), whereas at halftime starts measurement error variances stand out, being almost four times larger than at other occasions ($\hat\theta_{\text{HR},\text{init}}/\hat\theta_{\text{HR}}$).
\par

The expected football fever across the match is determined by the time point and the estimated expectations of the trend components, i.\,e., $\hat\mu_{I_{\text{FF}}}, \hat\mu_{S_1}$ and $\hat\mu_{S_2}$.
As heart rate is the marker indicator, the expectation of football fever closely resembles that indicator, peaking at kick-off ($\hat\mu_{I_{\text{FF}}}=\IntFF$) and linearly declining between $t=7$ and $t=24$ by 10\%.
Toward the end of the match, the expected football fever reverses its direction, rising with a faster pace than before (\IntSzwei\ vs.\ \IntSeins).
Still, we allowed fans to exhibit individual trends by modeling the trend as growth components with the following (lower triangular of the symmetric) variance-covariance matrix:
\begin{align*}
    \mathbb{V}\begin{pmatrix*}
    I_{\text{FF}} \\
    S_1 \\
    S_2
\end{pmatrix*} &= \begin{pmatrix*}[r]
    \vI & & \\
    \covISeins & \vSeins & \\
    \covISzwei & \covSeinsSzwei & \vSzwei
\end{pmatrix*} \\
\intertext{and}
\text{Cor}\begin{pmatrix}
    I_{\text{FF}} \\
    S_1 \\
    S_2
\end{pmatrix} &= \begin{pmatrix*}[r]
    1.00 & & \\
    -0.30 & 1.00 & \\
    0.66 & -0.11 & 1.00
\end{pmatrix*}.
\end{align*}

A large fraction of the variability in football fever is due to individual heterogeneity, reflected by the large variance of the random intercept 
($\hat\phi_{I_{\text{FF}}}=\vI$).
The variances of the slopes also show significant heterogeneity of fans regarding the initial decline (significance of $\phi_{S_1}$), but not regarding the late rally (large standard error of $\phi_{S_2}$).
However, these relationships between the trend components should be interpreted with caution, as only one covariance, namely $\phi_{I_{\text{FF}}, S_1}$, is statistically significant from zero.
\par
The autoregressive process $\{R_t\}_t$ is the second determinant of the common factor football fever, despite the growth components.
The carry-over effect of football fever is substantial across time points ($\hat\rho=\ARFF$). 
Individual deviations from the population trend thus persist and influence subsequent observations.
Analogous to $\hat\theta_{\text{HR},\text{init}}$, the variances of the autoregressive process are largest at the halftime starts ($\hat\phi_{R_\text{init}}>\hat\phi_{R}$).
\par 
We evaluate the fit of our time-dependent structural equation model by comparing it to a simpler baseline model, which imposes strict temporal measurement invariance (see Figure \ref{fig:app:baseline} in the Appendix for a visualization).
\begin{table}[ht]
    \centering
    \begin{tabularx}{\textwidth}{>{\raggedright\arraybackslash}X r r r r| r r r}
    \toprule
    &&&&& \multicolumn{3}{c}{\textbf{Corrected measures}}
    \\
    \textbf{Model} & \rotatebox[origin=l]{90}{\textbf{Pars}}  & \rotatebox[origin=l]{90}{\textbf{df}} \ \ & \rotatebox[origin=l]{90}{\textbf{AIC}} \ \ \  &  \rotatebox[origin=l]{90}{\textbf{SRMR}} \ \  & \rotatebox[origin=l]{90}{$\chi^2$} \ \ \ & \rotatebox[origin=l]{90}{\textbf{RMSEA}} \ \ &  \rotatebox[origin=l]{90}{\textbf{CFI}}\ \ \\
    \midrule
     Time-dependent SEM (see \ref{sec:model}) &20 & 2394 & 65207.2 & 0.079 & 2469.1 & 
     0.015 & 
     0.994
    \\[3ex]
     Time-invariant SEM (see \ref{fig:app:baseline})  &6 & 2408 & 76127.2 & 0.662 & 14035.6 & 0.187 & 0.000
    \\
    \bottomrule
    \end{tabularx}
    \caption{Number of free parameters (Pars), degrees of freedom (df), Akaike information criteria (AIC), standardized root mean residuals (SRMR), $\chi^2$ values, root mean squared errors of approximation (RMSEA), and comparative fit indices (CFI) of our time-dependent structural equation model and a time-invariant baseline model. The values are pooled across the 10 imputed datasets.}
    \label{tab:fitMeasures}
\end{table}
Table \ref{tab:fitMeasures} displays the fit of both models.
Our detailed structural equation model shows a clearly better fit ($\Delta \text{AIC} = 10920$), indicating that the incorporation of temporal dynamics is worthwhile despite the increase in model complexity by 14 additional parameters.
The standardized root mean squared residual of our model (0.079) is below the commonly cited threshold of 0.08, suggesting an acceptable model fit \citep{Schermelleh-Engel2003EvaluatingMeasures, Kline2023}.
\begin{figure}[!htbp]
    \centering
    \includegraphics[width=\linewidth]{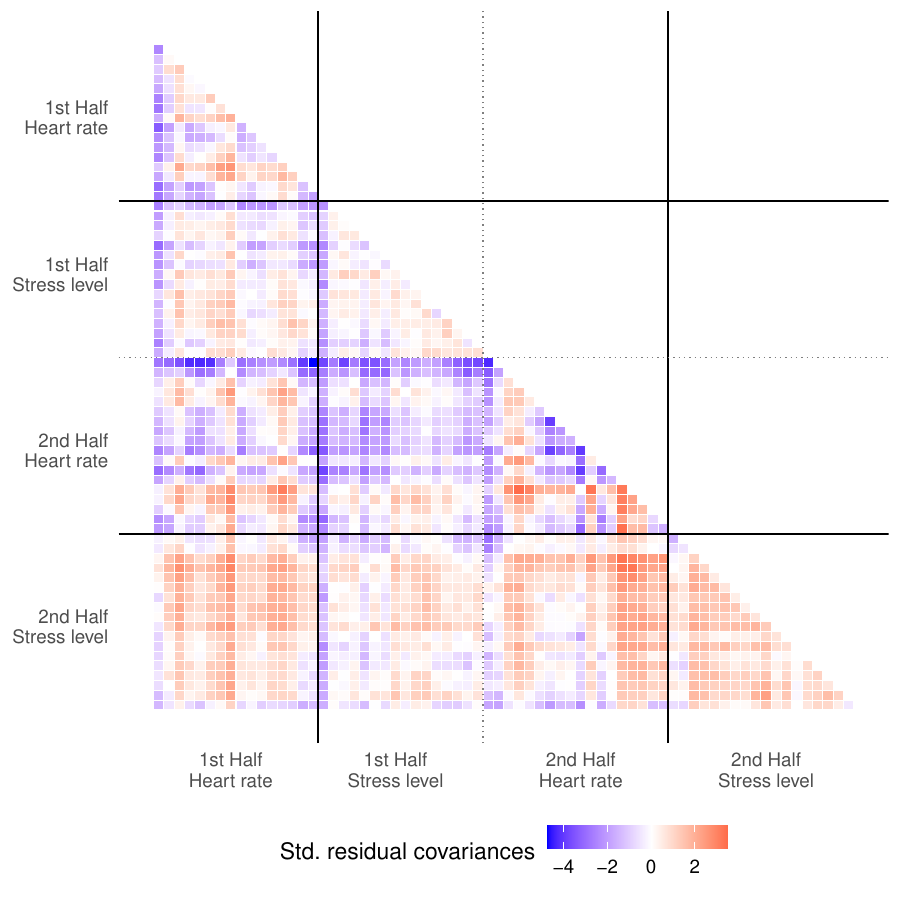}
    \caption{Heatmap of variances and covariances of standardized residuals, averaged across the 10 imputations, for all 34 measurement time points and both indicators. Solid black lines separate indicators and gray dotted lines halftimes.}
    \label{fig:stdResCov}
\end{figure}
For large models with many (co-)variances, however, literature suggests to additionally examine the individual residuals between model-implied and empirical (co-)variances of indicators \citep{Kline2023}.
Figure~\ref{fig:stdResCov} visualizes these standardized residuals between our sample and the estimated model. 
The matrix has a block-diagonal structure corresponding to the two halftimes and the two indicators.
Within-half residual covariances (both heart rate and stress level blocks) remain near zero, indicating that the trend adequately captures the temporal indicator dependencies within halves. 
Cross-half blocks exhibit strong negative residuals (second halftime heart rate with first halftime of both indicators), plus elevated second halftime heart rate residuals at halftime start ($t=17$) and Arminia Bielefeld goals ($t=29$).
%
\par
Many other fit measures, e.\,g., the comparative fit index, are based on the $\chi^2$ value.
However, this value and thus all derived fit measures are sensitive to the number of indicators and can be distorted in intensive time series data applications \citep[see][]{Norget2025EstimatingTutorial}. 
For this reason, we report corrected measures based on the $\chi^2$ correction in \citet{Yuan2015EmpiricalVariables}, which improves the type-I error performance of likelihood-ratio based tests in such settings.
As null model for the comparative fit index, we choose the time-invariant baseline model in Table \ref{tab:fitMeasures}.
The fit measures of the individual imputations are aggregated based on the pooling method in \citet{Chan2022MultipleTests}.
While our model's corrected root mean squared errors of approximation lies well below the typical threshold of 0.08 (at 0.015), that of the baseline model without time-dependencies exceeds it.
Likewise, the corrected comparative fit index of our model surpasses the threshold of 0.9 (at 0.994) \citep{Schermelleh-Engel2003EvaluatingMeasures}.
\par
Overall, our time-dependent structural equation model has a good model fit, which is substantially better than the time-invariant baseline, justifying additional parameters and enhancing our understanding of football fever dynamics.
\section{Discussion} \label{sec:discussion}
Based on the synchronous pathways in football fans' stress level and heart rate during the 2025 DFB Cup final, we assume an underlying common cause, football fever, driving the physiological measurements and representing fans' excitement and emotional arousal throughout a match.
Overall, the results obtained by building and estimating a detailed structural equation model largely support this hypothesis. 
The standardized factor loadings are large and positive, implying that high football fever is strongly associated with elevated heart rates and stress levels, and that both indicators reliably reflect this common factor. 
The reliability of heart rate aligns with physiological synchrony in fan crowds \citep{Baranowski-Pinto2022BeingSynchrony}, where arousal elevates collectively during matches.
In contrast, the pronounced between-fan heterogeneity in stress level -- captured by the random effects -- suggests more subjective appraisal of this shared arousal, which is consistent with previous literature finding individual differences in heart rate variability, a key stress factor, driving emotional regulation \citep{Krypotos2011IndividualStimuli}.
\par
Our model captures the temporal dynamics of football fever through latent growth components, revealing a V-shape trend for the average fans' emotional rollercoaster during the Cup final.
Presumably, this pattern reflects match-specific events, such as the early goals of VfB Stuttgart, leading to early frustration, and the late goals of Arminia Bielefeld, fueling hope.
However, fans appear to be heterogeneous in their temporal dynamics: For some, football fever remains almost constant, whereas others exhibit pronounced changes.
The covariances between the three latent growth components provide further insights in this regard. 
Fans with high initial football fever tend to be more strongly affected by critical match events than those with lower but more steady football fever.
These findings are in line with previous literature, linking fans hormonal reactions to their level of fandom \citep{vanderMeij2012TestosteroneFinal}.
Accordingly, we presume the strongly affected fans in our study to have a high level of fandom with a deep emotional bond, whereas the others are casual fans, who are affected less by the outcome.
Moreover, \citet{Gotz2020WhatCupMan} found group emotions, i.\,e., collectively felt emotions experienced live at the stadium, dominate fan identification based on club affiliation.
Consequently, the heterogeneity of fans in our sample may also stem from the means of sport consumption, with amplified football fever trajectories among those attending the match live at the Berlin stadium.
\par
Finally, the good overall fit of our model indicates that we can adequately capture the underlying common factor football fever and its dynamics.
The more detailed residual covariance matrix, on the other side, reveals systematic negative residuals associated with the start and halfway through the second halftime.
While the latter is likely due to the heterogeneity in fans' reactions to the chances and the goals of Arminia Bielefeld, the residuals concerning the start of the second halftime can have multiple reasons.
For instance, the residuals could be the result of a structural break from the temporal gap between halftimes or due to additional measurement non-invariance (e.\,g., rapid heart rate recovery while stress levels remain persistently elevated).
\section{Conclusion}  \label{sec:conclusion}
Our study demonstrates that we can adequately model football fever, defined as a common factor capturing fans' excitement and emotional arousal during football matches, is reliably reflected through physiological indicators such as heart rate and stress level and adequately modeled by our structural equation model. 
We identify a characteristic V-shape trend for football fever with latent growth components, reflecting match events alongside substantial between-fan heterogeneity.
The intensive longitudinal design of our wearable technology data advances beyond cross-sectional fan surveys by capturing current emotional dynamics throughout a high-stakes football match.
Understanding football fever dynamics is of multi-disciplinary interest.
Economically, it can inform real-time and fan-specific communication (peak engagement timing, event-aligned emotional highs) and fan segmentation. 
In addition, football fever affects vital parameters, making it an interesting factor for health benefits \citep[e.\,g.][]{Pringle2004CanMen}, and risks, e.\,g., cardiovascular events \citep{Puche2022SoccerEvents}.
Methodologically, our integration of wearable technology, multiple imputation, and time-dependent SEM, offers a framework for studying spectator emotions and physiological synchrony in real-world settings.
\par
There are limitations of our study to consider. 
First, the systematic cross-half residual covariances suggest unmodeled temporal dependencies or measurement non-invariance across halftime that our linear trend structure could not fully capture. 
Second, we lack additional information, such as level of fandom and viewing venue, which may comprise heterogeneous subpopulations. 
Third, our study design was restricted to fans with smartwatches from a particular manufacturer, who volunteered to share their personal data.
Since not all fans have such devices, our sample might be unrepresentative for the whole population of Arminia Bielefeld fans.
Moreover, the manufacturer uses a proprietary stress score, requiring further external validation against standardized measures \citep[e.\,g.][]{Rosenbach2025}.  
Finally, results reflect a single high-stakes Cup final in professional football, limiting generalizability to other contexts, e.\,g., amateur football.
\par
Future research should build upon our findings, for instance, by distinguishing live versus remote crowds, between low versus high fandom levels (i.\,e. enthusiastic and casual fans), or anti-synchronicity of fans of opposing teams. 
This work opens new avenues for quantifying the emotional and physiological impacts of sports spectatorship, bridging sports science, psychophysiology and consumer behavior.

\section*{Data availability}
Match-related data, including the minutes of goals and resulting scorelines, is publicly available. Requests for the data on vital parameters required to reproduce this study should be directed to the corresponding author.

\section*{Funding}
Our research was supported by the German Research Foundation (DFG) [RTG 2865/1 - 492988838], which we gratefully acknowledge.

\section*{Acknowledgments}
We sincerely thank all persons involved in the data collection, especially our colleague Christian Deutscher and the team of the Wissenswerkstadt Bielefeld for the fruitful collaboration in the joint data collection initiative.
\\
This manuscript was written with assistance from Perplexity’s large‑language‑model system for sentence refinement, formatting and code debugging. 
All model specifications, data analyses, interpretations, and scientific contributions originate exclusively from the authors. 
Core arguments and results represent the authors' original work.

\newpage
\printbibliography
\newpage
\appendix
\section{Baseline model}

\begin{figure}[ht]
    \centering
    \resizebox{\linewidth}{!}{
\begin{tikzpicture}[auto,node distance=.5cm,
    latent/.style={ellipse,draw,very thick,inner sep=0pt,minimum width = 2.5cm, fill=white,
	minimum height = 1.2cm,align=center},
    indicator/.style={rectangle,draw,very thick,inner sep=0pt,minimum width=20mm,minimum height=10mm},
    resid/.style={circle,draw,very thick,inner sep=0pt,minimum size=12mm,align=center, fill=white},
    paths/.style={->, very thick, >=stealth'},
    twopaths/.style={<->,very thick, >=stealth'}
]
\def\GTxPos{-7}
\def\leftX{-4}
\def\rightX{18}
\def\xFFdrei{15}
\def\xFFvier{20}

    \node at (0,0) [latent] (FF1) {$\text{football\_fever}_{1}$};
    \node at (5,0) [latent] (FF2) {$\text{football\_fever}_{2}$};
    \node at (\xFFdrei,0) [latent] (FF17) {$\text{football\_fever}_{17}$};
    \node at (\xFFvier,0) [latent] (FF18) {$\text{football\_fever}_{18}$};
    \node at (23.5,0) (FF19) {\huge$\dots$};

    \node [indicator, above left = 2cm and -1cm of FF1] (xFF21) {$\text{HR}_{1}$};
    \node [indicator, above right = 2cm and -1cm of FF1] (xFF11) {$\text{SL}_{1}$};
    
    \node [resid, above = 0.75cm of xFF11] (epsilon11) {$\varepsilon_{\text{SL}}$};
    \node [resid, above = 0.75cm of xFF21] (epsilon21) {$\varepsilon_{\text{HR}}$};

    \draw[paths] (FF1) -- node[sloped] {$\lambda_\text{SL}$} (xFF11);
    \draw[paths] (FF1) -- node[sloped] {$\lambda_\text{HR}$}(xFF21);
    \draw[paths] (epsilon11) -- (xFF11);
    \draw[paths] (epsilon21) -- (xFF21);

     \Cycle {epsilon11}{100}{15mm}[{node[yshift=9mm, xshift=-2mm]{$\theta_\text{SL}$}}]{6mm}{292};
     \Cycle {epsilon21}{100}{15mm}[{node[yshift=9mm, xshift=-2mm]{$\theta_{\text{HR}}$}}]{6mm}{292};

    \Cycle {FF1}{270}{15mm}[{node[yshift=-9mm, xshift=2mm]{$\phi_{\text{FF}}$}}]{6mm}{275};
    %
    %

    \node [indicator, above left= 2cm and -1cm of FF2] (xFF22) {$\text{HR}_{2}$};
    \node [indicator, above right = 2cm and -1cm of FF2] (xFF12) {$\text{SL}_{2}$};
    
    \node [resid, above = 0.75cm of xFF12] (epsilon12) {$\varepsilon_\text{SL}$};
    \node [resid, above = 0.75cm of xFF22] (epsilon22) {$\varepsilon_{\text{HR}}$};

    \draw[paths] (FF2) -- node[midway, sloped] {$\lambda_\text{SL}$} (xFF12);
    \draw[paths] (FF2) -- node[sloped] {$\lambda_\text{HR}$} (xFF22);
    \draw[paths] (epsilon12) -- (xFF12);
    \draw[paths] (epsilon22) -- (xFF22);

     \Cycle {epsilon22}{100}{15mm}[{node[yshift=9mm, xshift=-2mm]{$\theta_{\text{HR}}$}}]{6mm}{292};
     \Cycle {epsilon12}{100}{15mm}[{node[yshift=9mm, xshift=-2mm]{$\theta_\text{SL}$}}]{6mm}{292};

    \node at (8.5,0)  (FF3) {\huge$\dots$};

    \node at (8.5,5) (epsilon13) {\huge $\dots$};
    \Cycle {FF2}{270}{15mm}[{node[yshift=-9mm, xshift=2mm]{$\phi_{\text{FF}}$}}]{6mm}{275};

    %

    \node [indicator, above left= 2cm and -1cm of FF17] (xFF217) {$\text{HR}_{17}$};
    \node [indicator, above right = 2cm and -1cm of FF17] (xFF117) {$\text{SL}_{17}$};
    
    \node [resid, above = 0.75cm of xFF117] (epsilon117) {$\varepsilon_{\text{SL}}$};
    \node [resid, above = 0.75cm of xFF217] (epsilon217) {$\varepsilon_{\text{HR}}$};

    \draw[paths] (FF17) -- node[midway, sloped] {$\lambda_\text{SL}$} (xFF117);
    \draw[paths] (FF17) -- node[midway, sloped] {$\lambda_\text{HR}$} (xFF217);
    \draw[paths] (epsilon117) -- (xFF117);
    \draw[paths] (epsilon217) -- (xFF217);

     \Cycle {epsilon117}{100}{15mm}[{node[yshift=9mm, xshift=-2mm]{$\theta_\text{SL}$}}]{6mm}{292};
     \Cycle {epsilon217}{100}{15mm}[{node[yshift=9mm, xshift=-2mm]{$\theta_{\text{HR}}$}}]{6mm}{292};
     
    \Cycle {FF17}{270}{15mm}[{node[yshift=-9mm, xshift=2mm]{$\phi_{\text{FF}}$}}]{6mm}{275};

    %

    \node [indicator, above left= 2cm and -1cm of FF18] (xFF218) {$\text{HR}_{18}$};
    \node [indicator, above right = 2cm and -1cm of FF18] (xFF118) {$\text{SL}_{18}$};
    
    \node [resid, above = 0.75cm of xFF118] (epsilon118) {$\varepsilon_\text{SL}$};
    \node [resid, above = 0.75cm of xFF218] (epsilon218) {$\varepsilon_{\text{HR}}$};

    \draw[paths] (FF18) -- node[midway, sloped] {$\lambda_\text{SL}$} (xFF118);
    \draw[paths] (FF18) -- node[midway, sloped] {$\lambda_\text{HR}$} (xFF218);
    \draw[paths] (epsilon118) -- (xFF118);
    \draw[paths] (epsilon218) -- (xFF218);

     \Cycle {epsilon218}{100}{15mm}[{node[yshift=9mm, xshift=-2mm]{$\theta_{\text{HR}}$}}]{6mm}{292};
     \Cycle {epsilon118}{100}{15mm}[{node[yshift=9mm, xshift=-2mm]{$\theta_\text{SL}$}}]{6mm}{292};

    \node at (\xFFvier+3.5,5) (epsilon125) {\huge$\dots$};
    \Cycle {FF18}{270}{15mm}[{node[yshift=-9mm, xshift=2mm]{$\phi_{\text{FF}}$}}]{6mm}{275};
     \node[opacity=0.4, rotate = 90, scale=2.5, fill=white] at (11,3) (halftime) {\large Halftime break};

\end{tikzpicture}
}
    \caption{Baseline model.}
    \label{fig:app:baseline}
\end{figure}
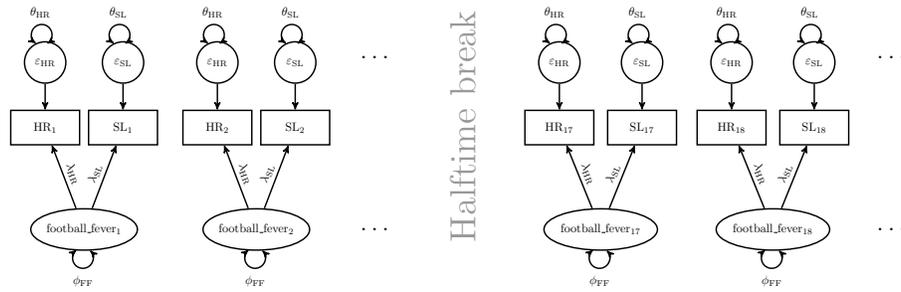

\section{Results for different imputation methods} \label{app:result_comparison}
\begin{table}[ht]
    \centering
    \setlength{\tabcolsep}{3pt}
    \hspace*{-0.05\textwidth}
    \begin{tabularx}{1.1\textwidth}{ll|rl|rl|rl|rl|rl|rl}
    \toprule
\multicolumn{2}{c}{\textbf{Parameter\ }} & 
\multicolumn{2}{c}{\textbf{R.\ forest}} & \multicolumn{2}{c}{\textbf{CART}} &
\multicolumn{2}{c}{\textbf{LASSO}} & \multicolumn{2}{c}{\textbf{Midastouch}}
& 
\multicolumn{2}{c}{\textbf{PMM}} & 
\multicolumn{2}{c}{\textbf{Normal}}\\
\midrule
    \parbox[t]{2mm}{\multirow{3}{*}{\rotatebox[origin=c]{90}{ICM}}} &  $\lambda_{\text{HR}}$ & 1.000 & & 1.000 &  & 1.000 &  & 1.000 &  & 1.000 &  & 1.000 &  \\ 
    & $\lambda_\text{SL}$ & 1.236 & $^{\clubsuit}$ & 1.161 & $^{\clubsuit}$ & 1.126 & $^{\clubsuit}$ & 1.123 & $^{\clubsuit}$ & 1.121 & $^{\clubsuit}$ & 1.131 & $^{\clubsuit}$ \\  
    & $\mu_{\text{SL}}$ & -40.782 & $^{\clubsuit}$ & -34.817 & $^{\clubsuit}$ & -33.02 & $^{\clubsuit}$ & -32.685 & $^{\clubsuit}$ & -32.505 & $^{\clubsuit}$ & -33.351 & $^{\clubsuit}$ \\ 
    \midrule
    \parbox[t]{2mm}{\multirow{6}{*}{\rotatebox[origin=c]{90}{Meas.\ error}}}
    & $\theta_{\text{HR},\text{init}}$ & 24.900 & $^{\clubsuit}$ & 22.457 & $^{\clubsuit}$ & 23.037 & $^{\clubsuit}$ & 22.754 & $^{\clubsuit}$ & 24.073 & $^{\clubsuit}$ & 22.033 & $^{\clubsuit}$ \\ 
    & $\theta_{\text{HR}}$ & 6.075 & $^{\clubsuit}$ & 4.439 & $^{\clubsuit}$ & 3.926 & $^{\clubsuit}$ & 4.238 & $^{\clubsuit}$ & 4.161 & $^{\clubsuit}$ & 3.677 & $^{\clubsuit}$ \\ 
    & $\theta_{\text{SL}}$ & 86.329 & $^{\clubsuit}$ & 95.297 & $^{\clubsuit}$ & 114.904 & $^{\clubsuit}$ & 109.406 & $^{\clubsuit}$ & 109.567 & $^{\clubsuit}$ & 114.33 & $^{\clubsuit}$ \\  
    & $\rho_{\text{SL}}$ & 0.895 & $^{\clubsuit}$ & 0.865 & $^{\clubsuit}$ & 0.647 & $^{\clubsuit}$ & 0.630 & $^{\clubsuit}$ & 0.637 & $^{\clubsuit}$ & 0.653 & $^{\clubsuit}$ \\ 
    & $\phi_{I_{\text{SL},\text{init}}}$ & 75.735 & $^{\clubsuit}$ & 77.179 & $^{\clubsuit}$ & 99.288 & $^{\clubsuit}$ & 105.078 & $^{\clubsuit}$ & 104.121 & $^{\clubsuit}$ & 98.265 & $^{\clubsuit}$ \\ 
    & $\phi_{I_{\text{SL},2}}$ & 28.060 & $^{\clubsuit}$ & 33.412 & $^{\clubsuit}$ & 45.145 & $^{\clubsuit}$ & 47.78 & $^{\clubsuit}$ & 47.751 & $^{\clubsuit}$ & 45.791 & $^{\clubsuit}$ \\ 
    \midrule
    \parbox[t]{2mm}{\multirow{9}{*}{\rotatebox[origin=c]{90}{Trend}}} & $\mu_{I_{\text{FF}}}$ & 89.046 & $^{\clubsuit}$ & 89.006 & $^{\clubsuit}$ & 88.996 & $^{\clubsuit}$ & 88.998 & $^{\clubsuit}$ & 88.998 & $^{\clubsuit}$ & 88.997 & $^{\clubsuit}$ \\ 
    & $\mu_{S_{1}}$ & -0.431 & $^{\clubsuit}$ & -0.428 & $^{\clubsuit}$ & -0.427 & $^{\clubsuit}$ & -0.428 & $^{\clubsuit}$ & -0.428 & $^{\clubsuit}$ & -0.428 & $^{\clubsuit}$ \\ 
    & $\mu_{S_{2}}$ & 0.633 & $^{\clubsuit}$ & 0.602 & $^{\clubsuit}$ & 0.603 & $^{\clubsuit}$ & 0.604 & $^{\clubsuit}$ & 0.603 & $^{\clubsuit}$ & 0.606 & $^{\clubsuit}$ \\ 
    & $\phi_{I_{\text{FF}}}$ & 164.518 & $^{\clubsuit}$ & 164.676 & $^{\clubsuit}$ & 164.851 & $^{\clubsuit}$ & 164.695 & $^{\clubsuit}$ & 164.727 & $^{\clubsuit}$ & 165.123 & $^{\clubsuit}$ \\ 
    & $\phi_{S_1}$ & 0.121 & $^{\clubsuit}$ & 0.114 & $^{\clubsuit}$ & 0.116 & $^{\clubsuit}$ & 0.114 & $^{\clubsuit}$ & 0.114 & $^{\clubsuit}$ & 0.117 & $^{\clubsuit}$ \\
    & $\phi_{S_2}$ & 0.114 & & 0.060 &  & 0.065 &  & 0.062 &  & 0.062 &  & 0.068 &  \\ 
    & $\phi_{I_{\text{FF}},S_1}$ & -1.428 & & -1.378 &  & -1.401 &  & -1.379 &  & -1.386 &  & -1.421 &  \\ 
    & $\phi_{I_{\text{FF}},S_2}$ & 2.505 & & 2.421 &  & 2.492 &  & 2.435 &  & 2.459 &  & 2.502 &  \\ 
    & $\phi_{S_1,S_2}$ & -0.023 & & -0.015 &  & -0.015 &  & -0.014 &  & -0.014 &  & -0.017 &  \\ 
    \midrule
    \parbox[t]{2mm}{\multirow{3}{*}{\rotatebox[origin=c]{90}{AR}}} & $\rho$ & 0.658 & $^{\clubsuit}$ & 0.673 & $^{\clubsuit}$ & 0.665 & $^{\clubsuit}$ & 0.670 & $^{\clubsuit}$ & 0.669 & $^{\clubsuit}$ & 0.660 & $^{\clubsuit}$ \\
    & $\phi_{R_\text{init}}$ & 82.646 & $^{\clubsuit}$ & 86.843 & $^{\clubsuit}$ & 84.731 & $^{\clubsuit}$ & 84.956 & $^{\clubsuit}$ & 84.427 & $^{\clubsuit}$ & 85.404 & $^{\clubsuit}$ \\ 
    & $\phi_{R}$ & 19.393 & $^{\clubsuit}$ & 19.323 & $^{\clubsuit}$ & 20.044 & $^{\clubsuit}$ & 19.635 & $^{\clubsuit}$ & 19.748 & $^{\clubsuit}$ & 20.391 & $^{\clubsuit}$ \\ 
    \bottomrule    
    \end{tabularx}
    \caption{Estimation of our time-dependent structural equation model: Parameter labels, estimates (pooled across 10 imputed datasets), and significance codes (based on the Bonferroni correction for 20 tests with threshold $\alpha^{\text{Bonf}}=0.0025$, indicated by black clubs~$^\clubsuit$). The column 'Random forest' contains the estimates from the original Table \ref{tab:estimates}.}
    \label{tab:estimates_comparison}
\end{table}

\begin{table}[ht]
\renewcommand{\arraystretch}{1.4}
    \centering
    \begin{tabularx}{\textwidth}{>{\raggedright\arraybackslash}X l r r r r| r r r}
    \toprule
    &&&&&& \multicolumn{3}{c}{\textbf{Corrected measures}}
    \\
    \rotatebox[origin=l]{90}{\textbf{Model}} & \rotatebox[origin=l]{90}{\textbf{Imputation}}& \rotatebox[origin=l]{90}{\textbf{Pars}}  & \rotatebox[origin=l]{90}{\textbf{df}} \ \ & \rotatebox[origin=l]{90}{\textbf{AIC}} \ \ \  &  \rotatebox[origin=l]{90}{\textbf{SRMR}} \ \  & \rotatebox[origin=l]{90}{$\chi^2$} \ \ \ & \rotatebox[origin=l]{90}{\textbf{RMSEA}} \ \ &  \rotatebox[origin=l]{90}{\textbf{CFI}}\ \ \\
    \midrule
     \multirow{6}*{\rotatebox[origin=l]{90}{Time-dependent SEM}} & R.\ forest &20 & 2394 & 65207.2 & 0.079 & 2469.1 & 0.015 & 0.994
    \\
     & CART &20 & 2394 & 64700.3 &    0.086 & 2706.7 & 0.031 & 0.974 
    \\
    & LASSO & 20 & 2394 & 65528.2 & 0.089 & 2877.3 & 0.038 & 0.959 \\
    & Midastouch &  20 & 2394 & 65339.7 & 0.093 & 3280.6 & 0.052 & 0.928 \\
    & PMM & 20 & 2394 & 65349.6 & 0.092 & 3211.7 & 0.050 & 0.933\\
    & Normal & 20 & 2394 & 65502.0 & 0.091 & 2938.5 & 0.041 & 0.954 \\
    \midrule
     \multirow{6}*{\rotatebox[origin=l]{90}{Time-invariant SEM}} & R.\ forest &6 & 2408 & 76127.2 & 0.662 & 14035.6 & 0.187 & 0.000 \\ 
     & CART &6 & 2408 & 75744.9 & 0.658 & 14341.0 & 0.190 & 0.000
    \\
    & LASSO &  6 & 2408 & 76374.6 &    0.638 & 14313.0 & 0.190 & 0.000 \\
    & Midastouch &  6 & 2408 & 76352.0 & 0.640 & 14708.9 & 0.193 & 0.000 \\
     & PMM & 6 & 2408 & 76352.0 & 0.640 & 14685.1 & 0.193 & 0.000 \\
    & Normal & 6 & 2408 & 76353.8 & 0.639 & 14332.6 & 0.190 & 0.000 \\
    \bottomrule
    \end{tabularx}
    \caption{Number of free parameters (Pars), degrees of freedom (df), Akaike information criteria (AIC), standardized root mean residuals (SRMR), $\chi^2$ values, root mean squared errors of approximation (RMSEA), and comparative fit indices (CFI) of our time-dependent structural equation model and a time-invariant baseline model. The values are pooled across the 10 imputed datasets. 'Random forest' refers to the original fit from Table \ref{tab:fitMeasures}.}
    \label{tab:fitMeasures_comparison}
\end{table}
\end{document}